# An Efficient monochromatic Electron Source designed for Inverse Photoemission Spectroscopy

GENG Dong-Ping(耿东平)[1,2]  YANG Ying-Guo(杨迎国)[1]  LIU Shu-Hu(刘树虎)[2]  HONG Cai-Hao(洪才浩)[2;1)]  GAO Xing-Yu(高兴宇)[1;2)]

[1] Shanghai Institute of Applied Physics, Chinese Academy of Sciences, Shanghai 201800, China

[2] Institute of High Energy Physics, Chinese Academy of Sciences, Beijing 100049, China

**Abstract:** A design of an efficient monochromatic electron source for Inverse Photoemission Spectroscopy (IPES) apparatus is described. The electron source consists of a BaO cathode, a focus electrostatic lens, a hemispherical deflection monochromator (HDM), and a transfer electrostatic lens. The HDM adopts a "slit-in and slit-out" structure and the degradation of first-order focusing is corrected by two electrodes between the two hemispheres, which has been investigated by both analytical methods and electron-ray tracing simulations using SIMION program. Through the focus lens, the HDM, and the standard five-element transfer lens, an optimal energy resolution is estimated to be about 53 meV with a beam flux of 27μA. Pass energy (P.E.) of 10 eV and 5 eV are discussed, respectively.

**Key words:** beam flux, electrostatic lens, hemispherical deflection monochromator, fringing effect, energy resolution, SIMION

**PACS:** 41.85.Ne; 41.85Qg; 41.85.Ja

## 1. Introduction

    Photoemission spectroscopy provides a powerful tool for the characterization of occupied electronic states, however, it cannot detect those unoccupied. Those unoccupied states can be probed by two-photo photoemission (2PPE) or inverse photoemission spectroscopy (IPES); nevertheless, 2PPE covers only a small part of a typical Brillouin zone in spite of its high energy resolution [1]. To the contrary, the dispersion of unoccupied states can be studied in a wider energy range above Fermi level with sufficient accuracy by IPES. In IPES, an incident electron is ejected into an unoccupied state and creates a radiative transition to another unoccupied state above the Fermi level, working in an inverse mode compared with PES. The photon emitted in this transition needs to be detected. Consequently, an electron source and a photon detector are essential in the IPES measurements. As the inverse photoemission process has rather low cross sections, the photon detectors in IPES suffer from quite low counting rates [2]. Therefore, special efforts should be dedicated to increase the intensity of the electron sources to make up the low counting rates of the outgoing photons [3]. To achieve high emission, the cathode of the electron sources often is made of material with low work function. However, the space charge effects become very severe for the low kinetic energy electrons used in IPES, which prevents the beam intensity becoming too high. On the other hand, as an important parameter determining the performance of the IPES apparatus, the energy resolution is determined by two factors: the energy spread of the electron source and the bandpass of the photon detector. Relatively narrow bandpass of the photon detector using Geiger-Müller counter can be achieved with various combinations of entrance windows and filled gases [4-6]. Thus, the energy spread of the electron source often becomes the key factor determining the overall energy resolution of the IPES apparatus [7].

* Supported by National Natural Science Foundation of China (Grant Nos. 11175239), One Hundred Person Project of the Chinese Academy of Sciences, and Instrument design and development project of CAS: Spin resolved Inverse-PES system.

1) E-mail: hongch@ihep.ac.cn

2) E-mail: gaoxy@sinap.ac.cn

In this article, we describe a design of monochromatic electron source with high efficiency for IPES. The whole system was analyzed and simulated by using electron-ray tracing simulation program, SIMION. To overcome the space charge limitation at low electron kinetic energy (KE) range, electrons in the present design are first generated at higher energy and later decelerated to the low KE used. In addition, a slit structure is used due to its larger acceptance than an aperture to achieve high beam intensity. To reduce the energy spread of the electron source, a hemispherical deflection monochromator (HDM) was adopted.

## 2. Design of the Electron Gun

The first unit of the electron source in this design consists of a cathode and a focus electrostatic lens, which forms an electron gun used in most applications. To improve the efficiency of an electron source, a cathode with high emission efficiency is necessary. For thermal emission, a BaO cathode is usually used because of its low cost, low energy spread due to its low working temperature, and high emission efficiency due to its low work function. The BaO cathode works at a temperature of about 1100 K. The energy distribution of thermally emitted electrons from the BaO cathode has its maximum at $E_{max}=kT$ with a half-width of about 2.45 kT (k represents the Boltzmann constant) [8]. Therefore, the natural energy spread of the BaO cathode can be calculated to be about 232.5 meV. The energy spread of the electron beam will stay constant after passing through an electrostatic lens due to the fact that all electrons are accelerated or decelerated together. Note that the initial electron velocity is not zero and emitted electrons have an energy spread. In the following analysis and simulation, we assume an initial electron KE of 0.4 eV.

Several designs of the electron gun for IPES can be found in the literature [3, 9-11], however, there are lots to be done to fully meet both demands of intense emission and a narrow energy spread desired by IPES. The brightness of Pierce diode gun [9] at quite low energies is limited by space charge effects because of the relatively low electric fields at the cathode surface. The low-voltage and high-current electron gun designed by Peter and Edward does not suit the electron KE of nearly 10 eV. As electrons from BaO cathode are emitted to a wide range of directions, lens systems are always used to focus as many as possible electrons to the desired direction in sophisticated electron guns for IPES. In a practical lens system, the number of the parameters one wants to control independently actually determines the complexity of a lens system. Generally, following requirements are often needed for a lens system: a. fixed object and image positions; b. control of beam angle; c. fixed linear or angular magnification; d. alterable electron energy at the exit side for electrons of fixed energy at the entrance side. Usually, minimum two cylinder lens elements are necessary to fulfill any one requirement above with one extra element added to fulfill one more requirement [12]. In the present design for IPES, the object and the image positions need to be fixed with the linear magnification constant to control the electron beam angle at the exit side, which means that the requirements a, b, and c need to be fulfilled. Therefore, a four cylinder element focus electrostatic lens is needed. In this article, a gun design similar to that used by Stoffel and Johnson [3] is adopted as shown in Fig. 1. BaO cathode, element 1 and element 2 as a whole are called immersion objective lens, which mainly pre-focus the electron beam. According to Child law, the space charge limited current density is [13]

$$J = \frac{4\varepsilon_0}{9}\sqrt{2\frac{e}{m}}\left(\frac{V^{3/2}}{d^2}\right) \qquad (1),$$

where V is the anode-cathode potential difference, d is the distance between the anode and the cathode, e is the negation of the electric charge carried by a single electron, m is the mass of a single electron, and $\varepsilon_0$ is the permittivity of vacuum.

From equation 1, increasing $V_{1f}$ in Fig. 1 will reduce the space charge effect with more electrons attracted from the cathode. In Figure 1, $V_{4f}$ determines the electron KE at the exit side and $V_{3f}$ is the key parameter to focus the electron beam.

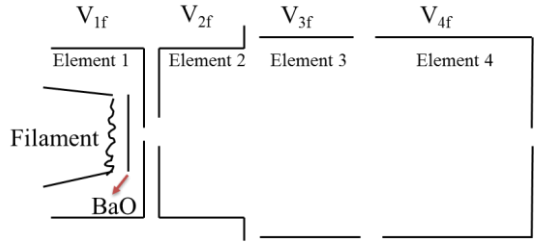

Fig. 1. The schematic diagram of the electron gun. $V_{1f}$, $V_{2f}$, $V_{3f}$ and $V_{4f}$ represent the voltages applied on element 1, 2, 3 and 4. Note that all the voltages applied are relative to the ground.

The simulation results using SIMION program are shown in Fig. 2. The electron gun can be operated with electron pass energies (P.E.) of 5 eV and 10 eV at the exit side. The characteristic lens diameters D1 and D2 are 8 mm and 12.5 mm, respectively. All the dimensional parameters of the lens are listed in Table 1. The lens tables (P.E. of 10 eV and 5 eV) are given in Table 2, respectively.

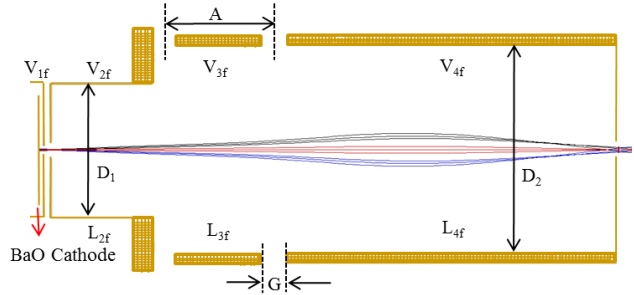

Fig. 2. Four cylinder elements simulation using SIMION program with P.E. of 10eV, where the green lines are the equipotential lines.

Table 1. The parameters of the four-element lens

| Parameter | A | G | $L_{2f}$ | $L_{3f}$ | $L_{4f}$ |
|---|---|---|---|---|---|
| Length (mm) | 6.25 | 1.25 | 12 | 5 | 18.75 |

Table 2. The lens table of P.E. 10eV and 5eV

| | BaO | $V_{1f}$ | $V_{2f}$ | $V_{3f}$ | $V_{4f}$ |
|---|---|---|---|---|---|
| P.E.=10eV | -10 | -6 | 200 | 500 | 0 |
| P.E.=5eV | -10 | -6 | 200 | 180 | -5 |

As Fig. 3 shows, Helmholtz-Lagrange Law [12] relates the linear magnification $M_L$ and the angular magnification $M_A$ of rays through an electrostatic lens to $E_o/E_i$, which can be expressed in the following equation,

$$(\frac{E_o}{E_i})^{\frac{1}{2}} = M_L M_A \qquad (2).$$

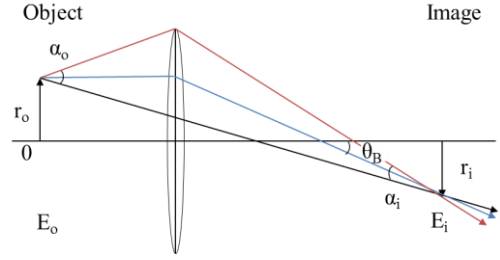

Fig. 3. The schematic drawing explains Helmholtz-Lagrange Law: $E_o$ and $E_i$ are the electron KE at the object and image position, respectively. Linear magnification is $M_L=r_i/r_o$, where $r_o$ and $r_i$ represent the displacements of the object and image, respectively. Angular magnification $M_A=\alpha_i/\alpha_o$, where $\alpha_o$ and $\alpha_i$ are the pencil angles at the object and image positions, respectively. Note that the pencil angle is the half of the beam angle $\theta_B$.

The beam angle $\theta_B$ (or the pencil angle $\alpha_i$) is a vital parameter to improve the energy resolution of HDM later and thus to improve the energy spread of the whole electron source. In the present design, the object position is located at BaO cathode with the image position at element 4 of the focus electrostatic lens. It can be seen that $E_o$=0.4 eV, $E_i$=10 eV for P. E. of 10 eV, or $E_i$=5 eV for P. E. of 5 eV, respectively. $\alpha_o$ is estimated to be about 90 ° based on the aperture structure of immersion objective lens. $r_0$ is designed to be 0.15 mm and $r_i$ ~1 mm for P.E. of 10 eV and $r_i$ ~1.1 mm for P.E. of 5 eV. The calculated values of $\alpha_i$ are compared with those obtained from SIMION simulation in Table 3. According to Table 3, the calculated $\alpha_i$ at the exit agrees well with the SIMION simulation results.

Table 3. The calculated $\alpha_i$ and the SIMION simulation result $\alpha_i$

| KE | $\alpha_i$(calculated) | $\alpha_i$(SIMION) |
|---|---|---|
| 10eV | 2.7 ° | 2.94 ° |
| 5eV | 3.47 ° | 3.4 ° |

Based on the above design, an electron gun was built and its beam currents were measured as functions of $V_{2f}$ for P. E. of 5eV and 10eV, which are reported in Fig. 4. From Fig. 4, it can be seen that $I_{2f} =I_{1f} -I_{2f}′$ and $I_{2f}$ reaches 200 μA when $V_{2f}$=200 V, where $I_{1f}$ ,$I_{2f}′$, and $I_{2f}$ represent the current passing element 1, that flowing through element 2, and that passing element 2, respectively. After $V_{2f}$ reaches 200 V, $I_{2f}$ does not increase much with increasing $V_{2f}$. Note that in P.E. of 10 eV, the beam current flux is higher than that in P.E. of 5 eV because higher anode voltage will extract more electrons from the cathode. In the experimental setup, there is a slit (as shown in Fig. 5) at the exit of element 4 to select only a part of the electrons through. Considering the electron beam image with a diameter of about 1mm at the exit of element 4, about 64% of the beam current can pass through this slit, which means that a beam flux of nearly 128 μA can be obtained with $V_{2f}$=200 V.

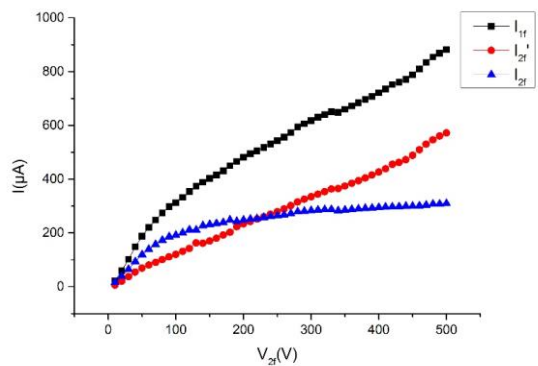

Fig. 4. Beam currents measured as functions of $V_{2f}$, where $V_{1f}= -6$ V, $V_{2f}=200$ V with a cathode negative bias voltage of 10 V. $I_{1f}$ represents the current passing element 1; $I_{2f}'$: the current flowing through element 2; $I_{2f}$: the current passing element 2.

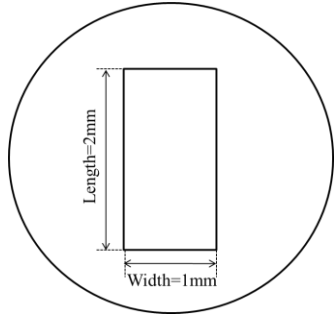

Fig. 5. The slit behind element 4

# 3. Design of the Hemispherical Deflection Monochromator

## 3.1 The principle of a Hemispherical Deflection Monochromator

As discussed above, the energy spread of the electron beam out of the focus lens in the present setup remains about 232.5meV. To reduce the energy distribution width, an HDM will be adopted. As shown in Fig. 6, if $E_0$ is the P.E. of electrons traveling along the orbit with a radius $R_0 = (R_1+R_2)/2$, then the voltages on the inner and outer hemispheres, $V_1$ and $V_2$, are given by

$$V_{1,2} = V_0 \left(\frac{2R_0}{R_{1,2}} - 1\right) \qquad (3).$$

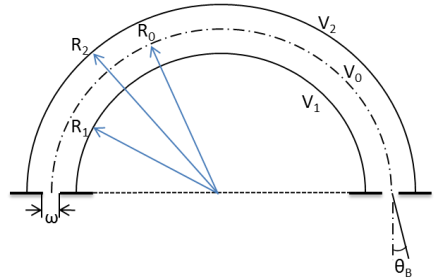

Fig. 6. A hemispherical deflection monochromator

With $R_1=50$ mm and $R_2=70$ mm, it can be obtained that $V_1=7$ V and $V_2=3.5714$ V for P.E.=5 eV, whereas that $V_1=14$ V and $V_2=7.1429$ V for P.E.=10 eV. The energy resolution or full width at half maximum (FWHM) of the electron with P.E. passing HDM can be given by [14]

$$\frac{\Delta E_{(FWHM)}}{P.E.} \approx \frac{0.86\omega}{2R_0} + 0.25\theta_B^2 \qquad (4)$$

In formula 4, $\omega$ and $\theta_B$ stand for the entrance slit width and the angle of the emitted electron beam divergency from the HDM as shwon in Fig. 6. The formula given above works when $\omega/R_0 \leq 0.1$, $\alpha_{max}/\alpha_0 \leq 2$, and $\alpha_{max} < 0.24$, where $\alpha_0=(\omega/4R_0)^{1/2}$. The entrance and exit slit width $\omega$ is 1 mm and the incidence beam angle is expected to be 5.9° for 10eV and 6.8° for 5 eV, respectively, after the focus electrostatic lens discussed in section 2. Accordingly, the energy resolution is about 53 meV for

P.E.=5 eV and 98 meV for P.E.=10 eV, respectively. Because the energy spread of electrons emitted from BaO cathod is 232.5 meV, the transmisssion of the HDA can be estimated to be about 0.212 for P.E.=10 eV and 0.381 for P.E.=10 eV. Thus, an eletron beam flux can reach 27 μA for P.E.=5 eV and 49 μA for P.E.=10 eV after the exit slit of the HDA. It can be seen that for P.E. of 5eV, a higher energy resolution (a smaller energy spread) can be achieved but with a lower electron beam flux. Therefore, a compromise must be made between the beam flux and the energy resolution. In most cases, an aperture structure is often adopted at both the entrance and exit of an HDM due to its simplicity. However, they are not suitable to achieve high beam flux due to that many electrons are blocked away. As a main advantage over Cylindrical Mirror Analyzer and other dispersion type energy monochromators, HDM can realize two-dimensional focusing to achieve high transmission with the same energy resolution. To take such an advantage, a rectangular slit structure shown in Fig. 5 was adopted at both entrance and exit with the one at the entrance serves as the exit slit of electron gun described previously. The overall setup is thus called "slit-in and slit-out" structure. Given the aperture diameter and the slit width to be the same 1 mm, the area of the slit in Fig. 5 can be calculated to be 2 mm$^2$ and that of the aperture is 0.8 mm$^2$. Obviously, the slit structure with a larger area is more efficient to transmit electrons than the aperture one [15]. From formula 4, ω is a key factor determining energy resolution in an HDM and a larger ω means a worse energy resolution. Using a larger aperture diameter to have higher electron transmission for the aperture structure, it is inevitable to enlarge the electron energy width. However, this problem can be solved when a slit structure shown in Fig. 5 is adopted. The length can be increased with more electrons transmitted but with the width kept constant without energy width being increased.

### 3.2 The correction of fringing effect

In practice, an HDM has fringing fields at both entrance and exit slits, which means that strong leakage fields inside and outside the hemispheres lead to field distortions, as shown in Fig. 7 (a). These fields obviously will hinder the performance of the HDM. Intense efforts have been dedicated to eliminate the fringing field effects and various correction schemes have been employed, such as Herzog plate, Jost electrodes, tilted input beam axis, and multiple rings or strips [16]. However, all these schemes have their own shortcomings, such as low effectiveness, complex fabrication, and difficult optimization [17].

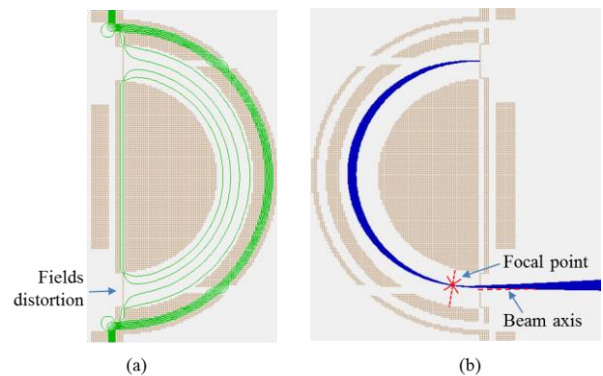

Fig. 7. (a) The SIMION simulation of an HDM when P.E.=10eV. The green lines denotes the equipotential lines. (b) Degradation of first-order focusing of an HDM when P.E.=10eV.

The simulation results of an HDM using SIMION program are shown in Fig. 7. In Fig. 7 (a), electric field distortions are present at both the entrance and exit slits. Owing to the distortions, the degradation of the first-order focusing is shown in Fig. 7 (b), in which the focal point (the red cross) is not located at the exit slit and the beam axis is shifted sideways. This can lead to transmission losses and a worse energy resolution. Here, a delicate and simple solution was adopted with two more electrodes added inside the space between the two hemispheres near both the entrance and exit, respectively, as shown in Fig. 8. By tuning the voltages of the two electrodes, the degradation of the first-order focusing can be corrected perfectly. From Fig. 8, it can be seen that the electron beam focus at the exit slit perfectly without beam axis angle shifted. The voltages applied on the five electrodes in Figure 8, which are optimized from SIMION program simulations, are listed in Table 4 for P.E. of 10 eV and 5 eV, respectively.

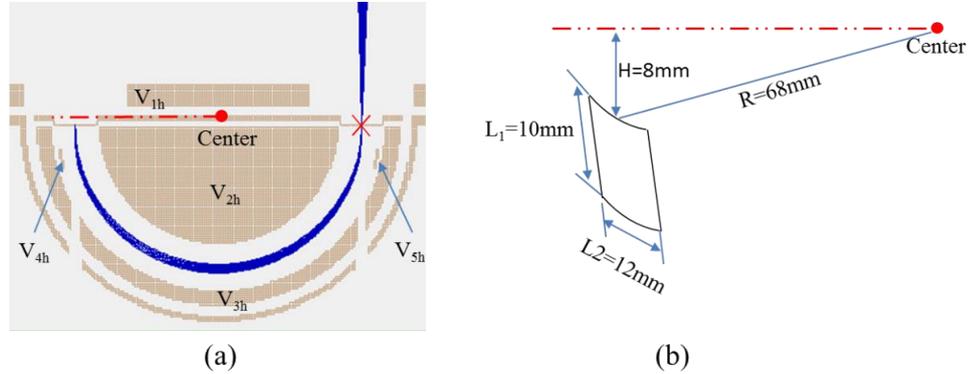

Fig. 8. (a) The HDM correction scheme with two electrodes added symmetrically at both sides (P.E.=10eV); (b)The detailed layout of the electrode structure at the entrance with its position between the hemispheres. The electrode is a part cut from a sphere (concentric with the two hemispheres) with a thickness of 0.1mm and a radius R of 68mm.

Table 4. The voltages of the five electrodes for P.E. 10eV and 5eV

|  | $V_{1h}$ | $V_{2h}$ | $V_{3h}$ | $V_{4h}$ | $V_{5h}$ |
|---|---|---|---|---|---|
| P.E.=10eV | 0 | 4 | -2.73 | -3.5 | -3.5 |
| P.E.=5eV | -5 | -3.055 | -6.4 | -6.75 | -6.6 |

## 4. Design of the transfer electrostatic lens

The ability to tune KE is essential for IPES. The electrons out of the HDM need to be accelerated or decelerated to tune their KE to probe different electronic structures at different energy levels above the Fermi level of the materials. Therefore, a transfer electrostatic lens have to be mounted after the exit slit of the HDM, which allows the electrons (at the constant 5eV or 10eV KE) to be tuned from 5eV to 20eV (the general scan KE range in IPES) continuously. To achieve high efficiency in IPES, the electrons of variable KE should be focused well on the samples after this lens.

Here, a standard five-element zoom lens is selected. A schematic view of this lens with its parameters is shown in Fig. 9. The distance unit used in this lens, or its characteristic value, is the cylinder diameter D=20 mm. Here, A/D=0.5 and G/D=0.1. The other distances of the lens are given in Table 5. The KE of ejected electrons from this lens is determined by the difference between $V_{5t}$ and $V_{1t}$. In this design, $V_{3t}/V_{1t}=V_{5t}/V_{3t}$. $V_{2t}$ and $V_{4t}$ of the middle electrodes can be tuned to keep the image position (i.e. the sample position) constant [18]. It should be noted that the HDM supporting plate ($V_{1h}$), element 4 of the focus lens ($V_{4f}$), and the first element of transfer lens ($V_{1t}$) are in electrical contact. Therefore, $V_{1t}=V_{1h}=V_{4f}$. In the five-element transfer lens in this paper, the KE of incoming electrons is denoted as $KE_0$ and that of outgoing electrons as $KE_1$. With the known $KE_0$, $KE_1$ and $V_{1t}$, $V_{5t}$ applied can be calculated as the following: $V_{5t} = KE_1/e - KE_0/e + V_{1t}$. The features of the standard five-element zoom lens (A/D=0.5) have been thoroughly studied about the relations of working distance and voltages of each element [18-21]. In practice, the lens needs to be constructed to find out the optimal conditions in the future experiments.

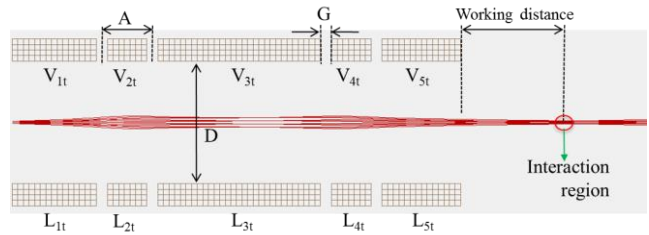

Fig. 9. The five-element transfer lens with its parameters: the upper V and lower L denote the voltage applied and the length of each element, respectively (P.E.=10eV).

Table 5. The lengths of the five-element zoom lens

| Element | $L_{1t}$ | $L_{2t}$ | $L_{3t}$ | $L_{4t}$ | $L_{5t}$ |
|---|---|---|---|---|---|
| Length(mm) | 25 | 10 | 50 | 10 | 25 |

Finally, electron-ray tracing simulations using SIMION program of the present HDM with fringing field correction and the five-element transfer lens as a whole are shown in Fig. 12. It is shown that electron beam through the HDM and the transfer lens can be well focused on the sample. Due to the fact that there is no energy spread and beam flux loss in the five-element transfer lens, the energy resolution and beam flux are estimated to be 98 meV and 49 µA for P.E.=10 eV, and 53 meV and 27 µA for P.E.=5 eV, respectively.

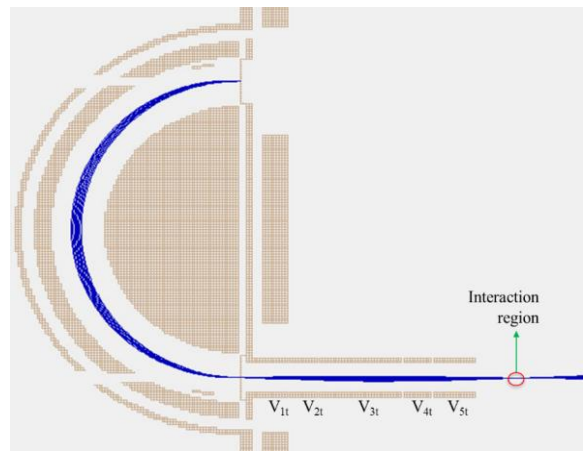

Fig. 12. The electron-ray tracing simulations of the HDM with fringing field correction and the transfer lens (P.E.=10eV).

## 5. Conclusion

In this paper, an efficient monochromatic electron source for IPES has been designed in details. The initial test shows that the beam flux can reach 128 µA at the exit slit of the focus electrostatic lens. A specially designed HDM is adopted after the focus electrostatic lens to reduce the energy spread of the electron beam. Adding two extra electrodes between the hemispheres perfectly solves the degradation of the first-order focusing after the HDM. The energy spread of the electron beam after the HDM is calculated to be 98 meV and 53 meV for 10 eV and 5 eV P.E., respectively. A "slit-in and slit-out" structure is adopted in favor of a high beam flux with a good energy resolution. A standard five-element zoom lens is then adopted after the HDM to tune the KE of the electrons emitted. The beam flux of the whole setup is estimated to be 49 μA for P.E.=10 eV, and 27 μA for P.E.=5 eV with an energy spread of 98 meV and 53 meV, respectively. The whole design makes the electron source possible to achieve high beam flux with narrow energy spread, which is ideal for IPES to study electronic states with high energy resolution.